\documentclass[twocolumn,floatfix,superscriptaddress,aps,prl]{revtex4-2}
\pdfoutput=1
\usepackage[utf8]{inputenc}
\usepackage[english]{babel}
\usepackage[T1]{fontenc}
\usepackage{amsmath}
\usepackage[normalem]{ulem}

\usepackage{graphicx}
\usepackage{amsmath,mathtools}
\usepackage{amssymb}
\usepackage{bm}
\usepackage{color}
\usepackage{bbold}
\usepackage{mathtools}

\DeclarePairedDelimiterX\braket[2]{\langle}{\rangle}{#1\,\delimsize\vert\,\mathopen{}#2}

\RequirePackage[
   hyperindex,colorlinks,bookmarksnumbered,
   plainpages=true,pdfstartview=FitH]{hyperref}
\hypersetup{linkcolor=blue,urlcolor=blue,citecolor=blue}
\usepackage{tikz}
\usepackage{quantikz}
\usepackage{lipsum}

\begin{document}

\title{Thermalization in a closed quantum system from randomized dynamics
}

\author{Nikolay V. Gnezdilov}
\email{nikolay.gnezdilov@dartmouth.edu}
\affiliation{Department of Physics and Astronomy, Dartmouth College, Hanover, New Hampshire 03755, USA}

\author{Andrei I. Pavlov}
\email{andrei.pavlov@kit.edu}
\affiliation{TKM, Karlsruhe Institute of Technology, 76131 Karlsruhe, Germany}

\begin{abstract}
The emergence of statistical mechanics from quantum dynamics is conventionally understood through quantum chaos, the Eigenstate Thermalization Hypothesis (ETH), and canonical typicality, which together explain the thermal behavior of local observables in quantum many-body systems. We provide a distinct mechanism in which, for a quantum many-body system under strong random perturbation, the chaotic behavior of the system's eigenvectors and a constraint on the total energy directly generate a canonical ensemble for the entire system. The canonical ensemble occurs as the first moment of the Porter-Thomas distribution of the system's chaotic eigenstates, and the constraint on the total energy sets the temperature in the ensemble. This yields canonically distributed expectation values without invoking ETH, a microcanonical ensemble, or a subsystem-bath partition for local and global observables. We numerically demonstrate the mechanism for the $1$D transverse-field Ising model by showing canonically distributed eigenstate projectors and spin-spin correlators with distinctive temperature dependence of the correlation length. An implementation of this thermalization approach on a quantum computer can be utilized for thermal state preparation.
\end{abstract}

\maketitle

{\bf Introduction.} 
In statistical mechanics, computing physical observables at finite temperature assumes the presence of a thermal bath -- a larger environment connected to a system that sets the temperature of the canonical ensemble via energy exchange with the bath. In contrast, at the quantum mechanical level, we often consider closed systems that evolve under unitary dynamics. The process through which quantum unitary dynamics gives rise to thermal observables is known as thermalization and is widely understood within the Eigenstate Thermalization Hypothesis (ETH)~\cite{Deutsch1991, Srednicki1994,Rigol2008,DAlessio2016}. In a closed quantum system that undergoes thermalization, canonical averages emerge at the subsystem's scale, where observables are measured locally, while the rest of the system effectively acts as a bath taken in the thermodynamic limit~\cite{Tasaki1998, Popescu2006, Goldstein2006Canonical}. In practice, numerical~\cite{Santos2012Weak,Genway2012Thermalization,Mierzejewski2013Eigenvalue,Garrison2018,Seki2020Emergence} and experimental~\cite{Kaufman2016, Somhorst2023Quantum} studies indicate that for a finite-size system split into a subsystem and a bath, local observables can approach their canonical averages.

Thermalization in finite-size systems has recently become particularly relevant for quantum computation. Potential applications of quantum computation to physics and chemistry imply the ability of programmable quantum processors to simulate thermal behavior~\cite{Reiher2017Elucidating,Alexeev2021Quantum,Fauseweh2023Quantum,schleich2025chemicallymotivatedsimulationproblems}. However, the current and near-term quantum processors operate with systems far from the thermodynamic limit. Hence, generating thermal observables for quantum few-body systems is essential. The natural thermalization process has inspired several methods for preparing thermal states on quantum hardware by engineering system-bath interactions/dissipative dynamics in a simulated system~\cite{shtanko2023preparingthermalstatesnoiseless,chen2023quantumthermalstatepreparation,chen2025efficientexactnoncommutativequantum, ding2025efficientquantumgibbssamplers,hahn2025provablyefficientquantumthermal,lloyd2025quantumthermalstatepreparation,hahn2025efficientquantumgibbssampling,ding2025endtoendefficientquantumthermal,Chen2025Efficient, li2025dissipativequantumalgorithmsexcitedstate}.

In this paper, we establish that observables matching those from the canonical ensemble can be obtained for an entire finite quantum system without subsystem-bath partition. Rather than reserving part of the system for a bath, we take a statistical approach, introducing random interactions within the system. 
Considering a system under a weak random-matrix perturbation paves a natural way to fulfill ETH~\cite{Reimann2015Eigenstate,Nation2018Off,Reimann2021Refining,cipolloni2023gaussianfluctuationsequipartitionprinciple,riabov2024eigenstatethermalizationhypothesiswignertype}. The weakness of the perturbation restricts the states contributing to an observable to a small microcanonical window, so that, for a typical chaotic eigenstate, an expectation value of the observable matches its microcanonical average. The latter is equivalent to the canonical average for observables with support substantially less than the system’s size~\cite{Touchette2015Equivalence,han2019adaptiveestimationshannonentropy,Tasaki2018,Kuwahara2020Eigenstate,Kuwahara2020Gaussian}. We take a different route that allows us to generate canonically distributed expectation values directly from the chaotic properties of the system’s eigenstates for both local and global observables.

We consider a target system Hamiltonian perturbed by random intra-system interactions with zero mean and large variance. The zero mean imposes a constraint on the total energy, conserved throughout the evolution, to equal the energy of the initial state (on average), and the large variance pushes the problem into a strong-perturbation regime. The random interaction is chosen such that the total Hamiltonian belongs to the Random Matrix Theory (RMT) universality class, while respecting a constraint on the total energy. As a consequence of strong perturbation, all energy eigenstates contribute to the expectation value of an observable rather than those concentrated in a small microcanonical window. Upon averaging over realizations of random interactions, we find that eigenstate expectation values are weighted by the Gibbs distribution with temperature set by the constraint on the total energy. We show that the Gibbs distribution emerges as the first moment of the Porter-Thomas (PT) distribution of overlaps of chaotic eigenstates with an initial state and occurs independently of an observable's support.
Unlike in conventional thermalization, where the canonical ensemble results from tracing out the bath degrees of freedom in a typical ``all-knowing’’ eigenstate of the system, in our approach, the Gibbs distribution of the expectation values is an emergent ensemble feature stemming from classical averaging.

For a finite spin chain used as our target system, we demonstrate our thermalization approach for both global (projectors) and local (spatially resolved correlation functions) observables: we derive many-body occupations of the energy eigenstates in the target system and spin-spin correlation functions. 
We show that within our approach, the global projectors become canonically distributed, unlike in conventional thermalization, and the same-time spatial correlation functions exhibit a temperature-dependent finite correlation length. The temperature is universally determined for all observables by the initial state's energy.

Overall, our approach relies on the universal properties of chaotic eigenstates for systems in the same RMT universality class, together with a global constraint on the average total energy. We therefore expect it to apply generally if these conditions are satisfied.

Implementing this approach in a few-qubit quantum simulation demonstrates that canonically distributed occupations of many-body eigenstates in a target system can be achieved in finite time~\cite{Perrin2025Dynamic}.

\begin{figure*}[t!!!]
\center
\includegraphics[width=0.7\columnwidth]{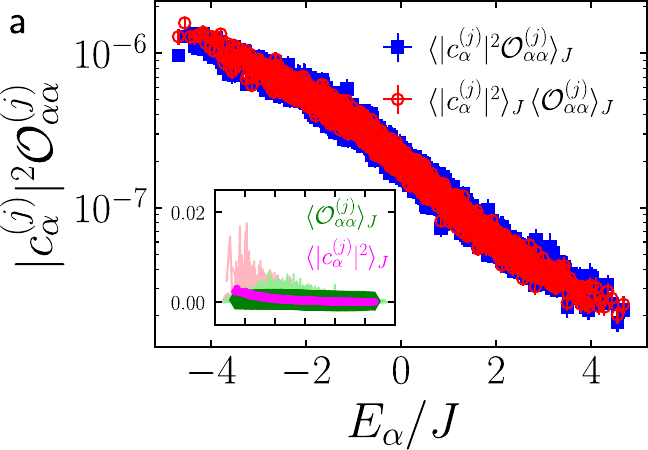}
\includegraphics[width=0.654\columnwidth]{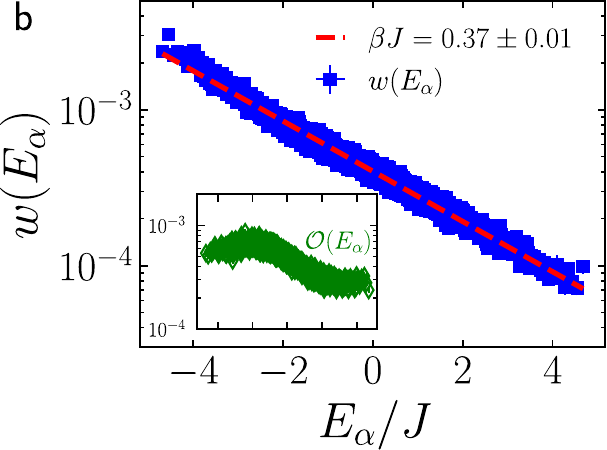}  
\includegraphics[width=0.628\columnwidth]{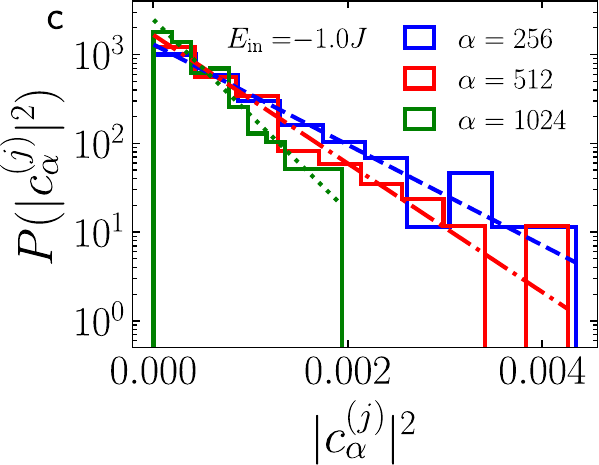} 
\caption{\small {\bf Thermalization from randomized dynamics}. 
Panel {\bf (a)} shows the eigenstate expectation value ${\cal O}^{(j)}_{\alpha\alpha}=\langle \psi^{(j)}_\alpha | {\cal O} |\psi^{(j)}_\alpha\rangle$ for ${\cal O}=|\varphi_\nu\rangle\langle \varphi_\nu|$ ($\nu = 500$) weighted by the eigenstate occupations in the initial state $|c^{(j)}_\alpha|^2=|\langle \psi^{(j)}_\alpha | \psi_{\rm in} \rangle|^2$, averaged over $M=200$ realizations of the system versus the many-body spectrum of $H$. For the horizontal axis, we use the mean spectrum of $H$: $E_\alpha=\langle E^{(j)}_\alpha\rangle_J$. The blue squares display $\langle|c^{(j)}_\alpha|^2 {\cal O}^{(j)}_{\alpha\alpha}\rangle_J$ and the red circles display $\langle|c^{(j)}_\alpha|^2\rangle_J \,\, \langle{\cal O} ^{(j)}_{\alpha\alpha}\rangle_J$: the two data sets lie right on top of each other. The inset shows $|c^{(j)}_\alpha|^2$ and ${\cal O}^{(j)}_{\alpha\alpha}$. The magenta dots denote the average values $\langle|c^{(j)}_\alpha|^2\rangle_J$, and the pink lines display every tenth realization of $|c^{(j)}_\alpha|^2$ that fluctuate from realization to realization. The green diamonds and the light green lines show the same for ${\cal O}^{(j)}_{\alpha\alpha}$. 
Panel {\bf (b)} shows the average eigenstate occupation probabilities $w(E_\alpha)=\langle|c^{(j)}_\alpha|^2\rangle_J$. 
The blue squares display $w(E_\alpha)$, and the dashed red line corresponds to the Gibbs distribution with the temperature $\beta^{-1}$ determined from Eq.~(\ref{beta}). In the inset, the green diamonds show the average eigenstate expectation value ${\cal O}(E_\alpha)=\langle{\cal O}^{(j)}_{\alpha\alpha}\rangle_J$. 
In panels {\bf (a)} and {\bf (b)}, the horizontal axis in the inset and in the main figure coincide.
In panel {\bf (c)}, the blue, red, and green histograms show the probability distributions for three eigenstates of $H$ sampled from $M$ realizations each. The correspondingly colored lines show a Porter-Thomas distribution, $P(p)=e^{-p/w(E_\alpha)}/w(E_\alpha)$, for each eigenstate. 
\label{fig:ETH}
}
\end{figure*}

{\bf The model.} 
We sample randomized unitary evolutions for the target Hamiltonian $H_0$, so that the state of the system evolves under the randomized Hamiltonian 
\begin{equation}
    H = H_0+{\cal V}^{(j)} \label{H}
\end{equation}
as $|\psi(t)\rangle = e^{-iHt}|\psi_{\rm in}\rangle$,
where $|\psi_{\rm in}\rangle$ is the initial state and ${\cal V}^{(j)}$ is the auxiliary random interaction between the degrees of freedom of $H_0$, realizations of which correspond to different samples of time evolution that we label by $j$. We choose the $1$D transverse-field Ising model (TFIM) with an open boundary condition as our target Hamiltonian: 
\begin{equation}
    H_0 = - h g \sum_{i=1}^{N} \sigma^x_i -h \sum_{i=1}^{N-1} \sigma^y_i \sigma^y_{i+1}, \label{H0}
\end{equation}
where $\sigma^x_i$ and $\sigma^y_i$ are the first and second Pauli matrices. We use the auxiliary interaction in the following form~\cite{Perrin2025Dynamic}
\begin{equation}
    {\cal V}^{(j)} = -\sum_{i_1>i_2}^N J^{(j)}_{i_1 i_2} \sigma_{i_1}^+ \prod^{i_1-1}_{k=i_2+1} \! \sigma_k^z \, \sigma_{i_2}^- + h.c. \label{V}
\end{equation}
The coupling constants $J^{(j)}_{i_1 i_2}$ are drawn independently from the Gaussian unitary ensemble (GUE) with zero mean $\langle J^{(j)}_{i_1 i_2} \rangle_J = 0$ and finite variance $\langle |J^{(j)}_{i_1 i_2}|^2\rangle_J = J^2/N$. 
An account of residual intra-system interactions among single-particle orbitals yields thermal orbital occupations in finite Fermi systems~\cite{Zelevinsky1996,Flambaum1997Distribution,Flambaum1997Statistical}, while considering weak random-matrix perturbations provides the conventional way to obtain thermal observables within the ETH paradigm~\cite{Reimann2015Eigenstate,Nation2018Off,Reimann2021Refining,cipolloni2023gaussianfluctuationsequipartitionprinciple,riabov2024eigenstatethermalizationhypothesiswignertype}. In contrast to the conventional thermalization approach, we choose the variance of the auxiliary interaction as the largest energy scale, pushing the problem into the strong-perturbation regime and mixing degrees of freedom across the many-body spectrum. 

Our model (\ref{H}) is ergodic, which implies that the system can achieve thermal equilibrium due to its own dynamics in the thermodynamic limit. The many-body energy spectra of ergodic systems exhibit Wigner–Dyson level statistics of random matrix theory (RMT). In the case of our model, the Hamiltonian~(\ref{H}) falls within the GUE class of RMT~\cite{Supplement}. If the system obeys RMT, for $N\to\infty$, we expect that a single realization of our model will be a typical one and suffice to produce thermal observables. In what follows, we demonstrate that averaging realizations of randomized evolutions yields canonical averages for the finite-size system without a microcanonical ensemble and a subsystem–bath partition. 

For further analysis, we choose the initial state to be a product state ordered along $\sigma^y$: $|\psi_{\rm in}\rangle =  \otimes_{k=1}^N \begin{pmatrix} i & 1 \end{pmatrix}_k^{\rm T}/\sqrt{2}$ \footnote{Later in the text we will also consider different initial states whose structure is discussed in the End Matter}, while the parameters of $H_0$ are $h=0.1 J$ and $g=1.5$, which correspond to the $\sigma^x$-ordered phase of the transverse Ising model at zero temperature. 
We set $N=11$ and define the $\alpha$th eigenstate $|\psi^{(j)}_\alpha\rangle$ and eigenenergy $E^{(j)}_\alpha$ of $H$ for its $j$th realization: $H |\psi^{(j)}_\alpha\rangle =E^{(j)}_\alpha |\psi^{(j)}_\alpha\rangle$. Similarly, we introduce the $\nu$th eigenstate $|\varphi_\nu\rangle$ and eigenenergy $\varepsilon_\nu$ of $H_0$: $H_0 |\varphi_\nu\rangle = \varepsilon_\nu |\varphi_\nu\rangle$.

{\bf Thermalization from randomized dynamics.} 
Let us compute the average expectation value of an observable $\cal O$:
\begin{equation}
    \langle\langle {\cal O}\rangle\rangle_J = \langle\langle \psi (t) | {\cal O} |\psi(t)\rangle\rangle_J, \label{O_def}
\end{equation}
where $\langle\ldots\rangle_J= \sum_{j=1}^M(\ldots)/M$ denotes the average over $M=200$ realizations of the time evolution of the state $|\psi(t)\rangle = e^{-iHt}|\psi_{\rm in}\rangle$. We decompose the initial state in terms of the eigenstates of the total Hamiltonian~(\ref{H}) for every realization of the latter: $|\psi_{\rm in}\rangle =\sum_\alpha c^{(j)}_\alpha |\psi^{(j)}_\alpha\rangle$. Then, we express the observable~(\ref{O_def}) as $\langle \langle {\cal O}\rangle\rangle_J \!=\!\! \sum_{\alpha_1\alpha_2}\!\langle c^{(j)*}_{\alpha_1}\!c^{(j)}_{\alpha_2}e^{it\Delta E^{(j)}_{\alpha_1\alpha_2}, 
    } {\cal O} ^{(j)}_{\alpha_1\alpha_2} \rangle_J$, where $\Delta E^{(j)}_{\alpha_1\alpha_2}\equiv E^{(j)}_{\alpha_1}-E^{(j)}_{\alpha_2}$ and ${\cal O}^{(j)}_{\alpha_1\alpha_2}\equiv \langle \psi^{(j)}_{\alpha_1} | {\cal O} |\psi^{(j)}_{\alpha_2}\rangle$. We perform a long-time average (denoted as $\overline{(\ldots)}$):
\begin{equation}
    \langle \langle\overline{\cal O}\rangle\rangle_J = \sum_\alpha \langle|c^{(j)}_\alpha|^2 {\cal O} ^{(j)}_{\alpha\alpha}\rangle_J, \label{O}
\end{equation}   
as a result of which, only diagonal-in-$\alpha$ terms contribute to Eq.~(\ref{O}), where $|c^{(j)}_\alpha|^2 = |\langle \psi^{(j)}_\alpha | \psi_{\rm in} \rangle|^2$ is the occupation probability of the $\alpha$th eigenstate of $H$ (for its $j$th realization) in the initial state and  ${\cal O}^{(j)}_{\alpha\alpha}=\langle \psi^{(j)}_\alpha | {\cal O} |\psi^{(j)}_\alpha\rangle$ is the corresponding eigenstate expectation value of the observable ${\cal O}$.

Averaging over randomized realizations of $H$ in Eq.~(\ref{O}) allows us to study statistics of $|c^{(j)}_\alpha|^2$ and ${\cal O} ^{(j)}_{\alpha\alpha}$ for every given $\alpha$.
In Fig.~\ref{fig:ETH}{\bf a}, we show that while $|c^{(j)}_\alpha|^2$ and ${\cal O} ^{(j)}_{\alpha\alpha}$ strongly fluctuate from realization to realization~\cite{Supplement}, their averaging over realizations can be performed independently, as they are statistically uncorrelated:
\begin{equation}
    \langle|c^{(j)}_\alpha|^2 {\cal O} ^{(j)}_{\alpha\alpha}\rangle_J = \langle|c^{(j)}_\alpha|^2\rangle_J \,\, \langle{\cal O} ^{(j)}_{\alpha\alpha}\rangle_J. \label{average}
\end{equation}
For the observable, we use the projector ${\cal O}=|\varphi_\nu\rangle\langle \varphi_\nu|$ onto the eigenstate of $H_0$ with $\nu = 500$. 
In Fig.~\ref{fig:ETH}{\bf a},{\bf b}, we plot $\langle|c^{(j)}_\alpha|^2 {\cal O} ^{(j)}_{\alpha\alpha}\rangle_J$, $\langle|c^{(j)}_\alpha|^2\rangle_J \,\, \langle{\cal O} ^{(j)}_{\alpha\alpha}\rangle_J$, $\langle|c^{(j)}_\alpha|^2\rangle_J$, and $\langle{\cal O}^{(j)}_{\alpha\alpha}\rangle_J$ versus the typical, i.e., mean, spectrum of $H$. Indeed, while the energy levels of $H$ can change slightly from realization to realization, the deviation from the average value $E_\alpha=\langle E^{(j)}_\alpha\rangle_J$ for each energy level is small, which is justified by the size of the horizontal error bars~\cite{Appendix}, which is smaller than the size of the marker. Hence, when factorized in the domain of realizations in Eq.~(\ref{average})~\footnote{The average of two uncorrelated, strongly fluctuating entities in Eq.~(\ref{average}) resembles the thermalization scenario $\#1$ in Ref.~\cite{Rigol2008}. In the latter scenario, the eigenstate occupations and eigenstate expectation values can exhibit strong fluctuations in a narrow energy window around the initial energy. However, these entities stay uncorrelated and average out independently in the chosen energy window. As a result, one obtains the observable consistent with the prediction of the microcanonical ensemble. The important distinction from this scenario is that, in our case, factorization of the average occurs in the domain of realizations rather than in the domain of energies.}, we consider the realization-averages $\langle|c^{(j)}_\alpha|^2\rangle_J$ and $\langle{\cal O} ^{(j)}_{\alpha\alpha}\rangle_J$ as functions of $E_\alpha$: $\langle|c^{(j)}_\alpha|^2\rangle_J=w(E_\alpha)$ and $\langle{\cal O} ^{(j)}_{\alpha\alpha}\rangle_J = {\cal O}(E_\alpha)$.

In Fig.~\ref{fig:ETH}{\bf b}, the average eigenstate occupations show finite overlap of the initial state with all energy eigenstates. This observation differs from conventional thermalization studies in which the broadening of the initial state in the energy space is set by the small strength of a perturbation that induces equilibration dynamics~\cite{Rigol2008}. This is expected since we consider our model in the strong perturbation regime, where the characteristic energy $J$ of the auxiliary interaction~(\ref{V}) is the largest energy scale in the problem, effectively extending the energy window around the initial state to the whole system. 
Thus, we account for all eigenstate expectation values ${\cal O}(E_\alpha)$, each weighted by $w(E_\alpha)$, to compute the expectation value of the observable
\begin{equation}
    \langle\langle \overline{\cal O}\rangle\rangle_J = \sum_\alpha w(E_\alpha) {\cal O}(E_\alpha), \label{O_th}
\end{equation}
where $\sum_\alpha w(E_\alpha)=1$. Consequently, as the energy window in Eq.~(\ref{O_th}) broadens to the entire system, the average equilibrium value of an observable does not match the microcanonical average expected in the weak perturbation regime widely applied in thermalization studies~\cite{Reimann2015Eigenstate,Nation2018Off,Reimann2021Refining,cipolloni2023gaussianfluctuationsequipartitionprinciple,riabov2024eigenstatethermalizationhypothesiswignertype}.

In the absence of the microcanonical ensemble in Eq.~(\ref{O_th}), we avoid infinite temperatures by a global constraint on the average total energy, $E_{\rm in} = \langle \psi_{\rm in}| H_0 |\psi_{\rm in}\rangle = \langle\langle H\rangle \rangle_J = \langle\langle \psi_{\rm in}| H|\psi_{\rm in}\rangle\rangle_J$, which is imposed by zero-mean coupling in the auxiliary interaction~(\ref{V}). Provided the constraint on the total energy, we examine whether the average occupations in the initial state, $w(E_\alpha)=\langle|\langle \psi^{(j)}_\alpha | \psi_{\rm in} \rangle|^2\rangle_J$, give rise to the canonical ensemble. In the latter case, the corresponding temperature should be found from the total energy conservation, so that the expectation value $\langle\langle H \rangle\rangle_J =\langle\langle \psi(t)| H |\psi(t)\rangle\rangle_J$ would agree with the prediction of the canonical ensemble~\cite{Srednicki1999}. In our model, this condition on temperature involves averaging over realizations.
Combining this with the energy constraint, the temperature in the canonical ensemble should be uniquely determined from 
\begin{equation}
   \begin{cases} \langle\langle H \rangle\rangle_J = \left\langle\dfrac{{\rm Tr} (e^{-\beta H}H)}{{\rm Tr}(e^{-\beta H})}\right\rangle_J \\ 
   E_{\rm in} = \langle\langle H \rangle\rangle_J \end{cases} \Rightarrow \quad \beta = \beta(E_{\rm in}), \label{beta}
\end{equation} 
where $\beta$ is the inverse temperature that explicitly depends on the choice of the initial state. 

\begin{figure}[t!!!]
\center 
\includegraphics[width=0.68\columnwidth]{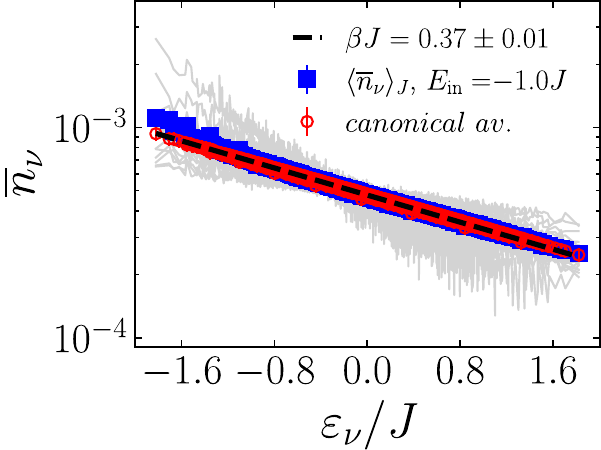}  
\caption{\small  {\bf Occupation probabilities of eigenstates of the TFIM} calculated by averaging randomized evolutions. 
The gray lines show the occupation probabilities for every tenth of the $200$ realizations. The blue squares are the average occupation probabilities found from Eq.~(\ref{O}). The red circles are occupation probabilities found from Eqs.~(\ref{O_th})-(\ref{Gibbs}).
The dashed black line shows occupations populated according to the Gibbs distribution with the temperature found from Eq.~(\ref{beta}). 
\label{fig:ns}
}
\end{figure}

In Fig.~\ref{fig:ETH}{\bf b}, we demonstrate that the average eigenstate occupation probabilities match the Gibbs distribution,
\begin{align}
    w(E_\alpha) = e^{-\beta E_\alpha}/{\cal Z}, \label{Gibbs}
\end{align}
where ${\cal Z} = \sum_\alpha e^{-\beta E_\alpha}$ and $\beta$ is determined from the condition~(\ref{beta}). 

The origin of the Gibbs distribution for the average eigenstate occupations $w(E_\alpha)=\overline{|\langle \psi^{(j)}_\alpha | \psi_{\rm in} \rangle|^2}$, where $|  \psi^{(j)}_\alpha \rangle$ is the eigenstate of the Hamiltonian in the GUE class of RMT, is rooted in the properties of random quantum states of ergodic systems~\cite{Brody1981RMT, Mirlin2000, Boixo2018} that can also support global constraints~\cite{Mark2024Maximum, mcginley2025scroogeensemblemanybodyquantum}. Within our approach, sampling many randomized realizations of the system allows us to draw the probability distribution for the eigenstate occupations $|c^{(j)}_\alpha|^2 = |\langle \psi^{(j)}_\alpha | \psi_{\rm in} \rangle|^2$ for each energy eigenstate (at fixed $\alpha$). In Fig.~\ref{fig:ETH}{\bf c}, we show that this probability distribution follows the Porter-Thomas distribution~\cite{PT1956}, $P(|c^{(j)}_\alpha|^2)=e^{-|c^{(j)}_\alpha|^2/w(E_\alpha)}/w(E_\alpha)$, characteristic of chaotic behavior~\cite{Brody1981RMT, Mirlin2000, Boixo2018}. Subsequently, averaging realizations of each eigenstate occupation probability and plotting the mean occupations versus the mean spectrum of $H$ leads to Fig.~\ref{fig:ETH}{\bf b} and Eq.~(\ref{Gibbs}). In this way, the Gibbs distribution emerges as the first moment of the PT distribution, with the temperature set by imposing a global constraint on the total energy. Changing the initial state~\cite{Appendix} accordingly changes the temperature in $w(E_\alpha)$, so that the eigenstate occupations depend only on $E_{\rm in}$ but not on the initial state's structure~\cite{Supplement}.

This direct statistical derivation of the canonical ensemble from the PT distribution, expected for the considered universality class, and the global constraint is different from the conventional thermalization, where, provided the system thermalizes in a microcanonical sense under a weak perturbation~\cite{Reimann2015Eigenstate,Nation2018Off,Reimann2021Refining,cipolloni2023gaussianfluctuationsequipartitionprinciple,riabov2024eigenstatethermalizationhypothesiswignertype}, the canonical ensemble occurs as a consequence of a subsystem-bath partition for local observables~~\cite{Touchette2015Equivalence,han2019adaptiveestimationshannonentropy,Tasaki2018,Kuwahara2020Eigenstate,Kuwahara2020Gaussian}. In our case, the canonical ensemble stems from the average eigenstate occupations $w(E_\alpha)$ at the level of Eq.~(\ref{O_th}) and does not rely on the choice of an observable and its support.

{\bf Observables.} 
To explicitly demonstrate that our approach produces canonically distributed global and local observables, we compute $\langle\langle\overline{\cal O}\rangle\rangle_J$ for global projectors and spatial spin-spin correlation functions.
We begin by computing the average expectation value for the projector ${\cal O}=|\varphi_\nu\rangle\langle \varphi_\nu|$ for all eigenstates $|\varphi_\nu\rangle$ of the Ising Hamiltonian $H_0$ from Eq.~(\ref{O}) and from Eqs.~(\ref{O_th})-(\ref{Gibbs}). 
As seen in Fig.~\ref{fig:ns}, occupation probabilities of the eigenstates of the TFIM calculated by averaging randomized dynamics, $\langle\overline{n}_\nu\rangle_J = \langle\overline{\langle \psi(t)|{\cal O} |\psi(t)\rangle}\rangle_J$, are thermal. The ensemble temperature is set by the energy-conservation constraint~(\ref{beta}), and changes for different initial states~\cite{Supplement}. For reference, we plot populations of the eigenstates of the Ising model according to the Gibbs distribution $e^{-\beta\varepsilon_\nu}/\sum_\nu e^{-\beta\varepsilon_\nu}$ with $\beta$ found from Eq.~(\ref{beta}). In contrast to ETH, which does not hold for global projectors~\cite{DAlessio2016}, we have shown that their expectation values agree with canonical predictions when averaged over chaotic dynamics under a global energy constraint.

\begin{figure*}[t!!!]
\center
\includegraphics[width=0.68\columnwidth]{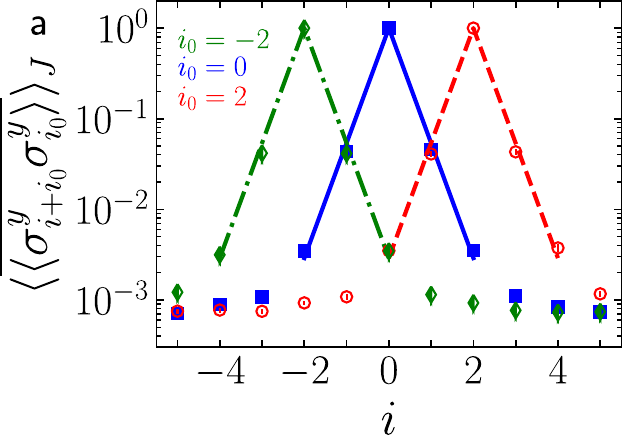} 
\includegraphics[width=0.68\columnwidth]{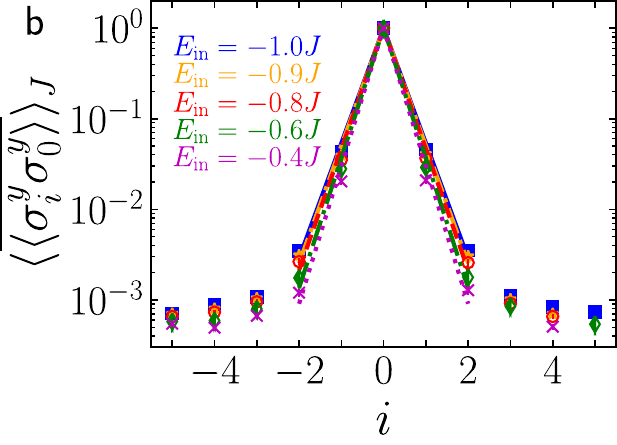} 
\includegraphics[width=0.644\columnwidth]{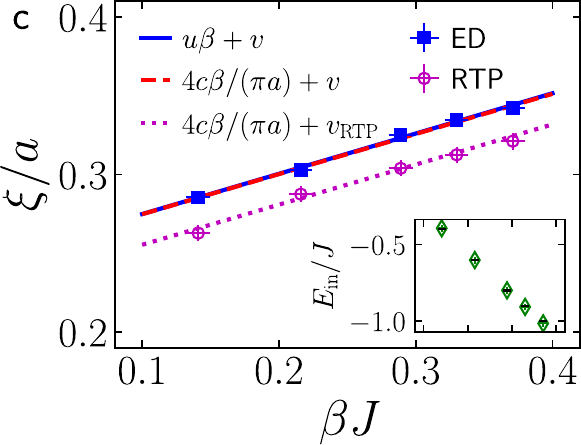} 
\caption{\small {\bf Correlation functions of the TFIM}. 
Panel {\bf (a)} shows the average same-time spin-spin correlators as functions of the lattice sites on a log scale. The blue squares represent the correlator positioned at $i_0=0$ and the red circles and green diamonds display the correlators placed at $i_0=2$ and $i_0=-2$. The correspondingly colored lines show the exponential decay with the rate determined from fitting the central correlator in the interval $i = [-2,2]$. In panel {\bf (b)}, the blue squares, red circles, green diamonds, and magenta crosses show correlators at $i_0=0$ for gradually increasing initial energies. The correspondingly colored lines show the fits performed at every initial energy $E_{\rm in}$.  
In panel {\bf (c)}, we plot the correlation length $\xi$ as a function of the inverse temperature $\beta$ determined by $E_{\rm in}$ (with small uncertainties in the total energy $\langle\langle H\rangle\rangle_J$ that stem from averaging over realizations). The blue squares show $\xi$ found from the fit in the central panel for the correlators computed using exact diagonalization (ED). The magenta circles show the same for the correlators computed with the real-time propagation (RTP). 
The solid blue line is the linear fit to the data, and the dashed red line shows predictions for $\xi$ in the thermodynamic limit up to the finite-size offset. In the inset, the green diamonds and black dashes display $\langle\langle H\rangle\rangle_J$ and $E_{\rm in}=\langle\psi_{\rm in} |H_0|\psi_{\rm in}\rangle$ as a function of $\beta$. The horizontal axis is the same as in the main figure. 
\label{fig:YY}
}
\end{figure*}

In our previous analysis, we established the properties of the eigenstates of $H$ and computed expectation values of the global projectors within our approach.
Both tasks require that eigenstates and eigenenergies of the system are known, which is generally not the case in quantum many-body physics. Instead, thermal behavior manifests itself in correlation functions that can be computed theoretically or measured experimentally~\cite{Trotzky2012}. For the transverse-field Ising model~(\ref{H0}), by comparing the model's parameters with the determined temperature ($\beta^{-1} \approx 2.7 J \gg 2h|1-g| = 0.3 J$), we expect the same-time spatial spin-spin correlation function, $\langle\sigma^y(x)\sigma^y(0)\rangle\sim e^{-|x|/\xi}$, to decay in the continuum limit with the linear-in-$\beta$ correlation length $\xi = 4 c \beta/\pi$~\cite{Sachdev2011}; $c = 2 h a$ and $a$ is the lattice constant. Having established that canonical averages can be derived for the relatively small isolated quantum system, we examine whether averaging randomized evolutions yields thermal behavior in the correlation functions.
To do so, we compute $\langle\overline{\langle\sigma^y_{i+i_0}\sigma^y_{i_0}\rangle}\rangle_J$ using Eq.~(\ref{O}) for the correlation functions centered at $i_0 = -2, 0, 2$. 

In Fig.~\ref{fig:YY}{\bf a}, the average correlation functions show a clear exponential decay over two sites on either side of their center. Then, the decay slows down and tends to saturate to a small value, which we attribute to the finite size of the system. To determine the decay rate, we fit the logarithm of the data for the centrally positioned correlator ($i_0=0$) by $-|x_i|/\xi$ in the interval $i = [-2, 2]$, with $\xi$ being a fitting parameter and $x_i = i a$. We compare the spatially displaced correlators ($i_0=-2,2$) and the decaying function $e^{-|x_{i\pm 2}|/\xi}$, where $\xi$ was found by fitting the central correlator. All correlation functions in the bulk of the system decay at the same rate and show good agreement with the fit. Since the initial energy determines the temperature~(\ref{beta}), we compute the correlation length as a function of $\beta$ by repeating the calculation for the central correlator for five initial states with distinct $E_{\rm in}$~\cite{Appendix}. Increasing the initial energy leads to faster decay of the correlator, as seen in Fig.~\ref{fig:YY}{\bf b}. In Fig.~\ref{fig:YY}{\bf c}, we show that increasing the initial energy raises the temperature and suppresses the correlation length. The correlation length is linear in the inverse temperature, as expected in statistical mechanics. We fit $\xi$ with a linear function $\xi = (u\beta +v)a$. We find an offset $\Delta\xi = v a = (0.250\pm0.004) a$ due to the non-zero lattice spacing in the finite-size system. However, the slope $u=(0.256\pm 0.016)J$ matches the prediction of statistical mechanics $4c/(\pi a)\approx 0.255 J$  within the error bar. 

One can avoid the costly exact diagonalization procedure and compute thermal observables from the real-time propagation by directly evolving the state in Eq.~(\ref{O_def}) and averaging over realizations~\footnote{In the Supplemental Material~\cite{Supplement}, we show that spin-spin correlation functions computed in the infinite-time limit, averaged over a finite time interval, and at a single time point exhibit the same characteristic decay.}. 
The average observables relax to the thermal equilibrium values in finite time $t\sim 1/J$~\cite{Perrin2025Dynamic, Supplement}. Using Eq.~(\ref{O_def}) for $Jt = 10, 20, 30, 40$, we repeat the calculation for the spin-spin correlation functions. After time-averaging the central correlator for each initial energy with the four time points, we apply the same fitting procedure as above to determine the correlation length. As seen in Fig.~\ref{fig:YY}{\bf c}, the computation recovers the temperature dependence of the correlation length derived earlier, up to a finite-size offset.

Above, we demonstrated that our approach does not differentiate between global and local observables and, in contrast with ETH, yields canonical predictions for both. Indeed, the chaotic character of individual many-body eigenstates manifests itself in the emergence of the Gibbs weights $w(E_\alpha)$ in Eq.~(\ref{O_th}) after averaging over realizations and does not rely on the choice of an observable. This allows randomized Hamiltonian dynamics to simulate thermal behavior beyond local or single-particle observables~\cite{Martin2023Equilibration,kiss2025neutrinothermalizationrandomizationquantum}.

{\bf Discussion.} 
In this work, we provided a way to obtain the canonical ensemble for a target system {\it without} the prerequisite of a microcanonical average and the following subsystem-bath partition, which are inherent in the context of ETH. To do so, we combine the target Hamiltonian with a random auxiliary interaction in the strong-perturbation regime so that the total Hamiltonian (i) belongs to the RMT universality class and (ii) respects a global constraint on the average total energy. Rather than tracing a larger part of the system in a typical pure state, we compute a classical average of evolved observables over realizations of the auxiliary interaction. As a result, canonical observables directly arise from the chaotic nature of individual many-body eigenstates. By sampling realizations of the auxiliary interaction, we find that the occupation probability of each many-body eigenstate in the initial state follows the PT distribution (expected for any system belonging to the same universality class). Averaging over realizations then leads to each eigenstate expectation value of the observable in Eq.~(\ref{O_th}) being weighted by the Gibbs distribution function~(\ref{Gibbs}) that appears as the first moment of the PT distribution. Consequently, average observables, including global projectors and many-body correlation functions, agree with their canonical averages even for a finite-size system. The temperature in the canonical ensemble is determined by a global constraint on the average total energy imposed by the initial state~(\ref{beta}). This result provides a direct derivation of the canonical ensemble from universal properties of chaotic eigenstates in the presence of a global constraint. Having shown the result's independence of the initial state's structure, we expect it to hold if conditions (i) and (ii) are met, irrespective of the microscopic details of the system.

The reliance of the approach on general properties of chaotic eigenstates and a global constraint is beneficial for applications to different (or larger) systems in classical and quantum simulations, which may require reducing the connectivity of the auxiliary interaction~\footnote{The auxiliary interaction with reduced connectivity will not couple all degrees of freedom in the system instantaneously, but spread through the system in finite time, which may increase the equilibration time.}. 
The ability of the approach to generate thermal observables from real-time evolution in finite time enables some quantum simulation methods~\cite{Perrin2025Dynamic} and classical algorithms, such as the time-dependent variational principle with matrix product states~\cite {Haegeman2011TDVP,Haegeman2016Unifying}, to produce canonical averages without partitioning the system into a subsystem and a bath.
Applying our thermalization procedure to a target system, we microscopically sample chaotic eigenstates respecting global constraints. Such sampling creates the possibility of generating random circuits that implement constrained Haar-random-like evolutions~\cite{Mark2024Maximum,mcginley2025scroogeensemblemanybodyquantum,mok2026naturestingyuniversalityscrooge} from quantum few-body Hamiltonian dynamics.

{\bf Acknowledgment.}
We thank Vanja Dunjko, Yuval Gefen, Archana Kamal, Maxim Olshanii, and Thibault Scoquart for stimulating discussions. N.V.G. thanks Alexander Shnirman for the hospitality during the visit to KIT. N.V.G. was supported by the Department of Physics and Astronomy, Dartmouth College. A.I.P. was supported by the German Ministry of Education and Research (BMBF) within the project QSolid (FKZ: 13N16151) and by the DFG Grant SH 81/8-1.

{\bf Data availability.} The numerical data produced in this study are available in an open repository~\cite{data}.

\bibliography{refs}

\appendix
\section{Appendix} \label{appendix}

{\bf Construction of initial states.} We additionally consider three initial product states constructed by spin flips (in $\sigma^y$ direction) of the initial state defined below Eq.~(\ref{V}) so that the initial states are non-degenerate in their initial energy. We use the following product states: $|\psi_{\rm in}\rangle =  \left(\otimes_{k=1}^{N-1} |\downarrow_y\rangle_k\right) \otimes |\uparrow_y\rangle_N$, $\left(\otimes_{k=1}^{N-2}|\downarrow_y\rangle_k\right) \otimes |\uparrow_y\rangle_{N-1}\otimes |\downarrow_y\rangle_{N}$, and $\left(\otimes_{k=1}^{N-3}|\downarrow_y\rangle_k\right) \otimes |\uparrow_y\rangle_{N-2}\otimes |\downarrow_y\rangle_{N-1}\otimes |\uparrow_y\rangle_{N}$, where $|\downarrow_y\rangle = \begin{pmatrix} i & 1 \end{pmatrix}^{\rm T}/\sqrt{2}$ and $|\uparrow_y\rangle = \begin{pmatrix} -i & 1 \end{pmatrix}^{\rm T}/\sqrt{2}$. The corresponding initial energies are $E_{\rm in}=-0.8J$, $-0.6 J$, and $-0.4 J$. 
We also consider a superposition initial state $|\psi_{\rm in}\rangle = \left[\otimes_{k=1}^{N} |\downarrow_y\rangle_k+ \left(\otimes_{k=1}^{N-1} |\downarrow_y\rangle_k\right) \otimes |\uparrow_y\rangle_N\right]/\sqrt{2}$ with $E_{\rm in}= -0.9 J$.

{\bf Error bars.} Fig.~\ref{fig:ETH}.  In panel {\bf (a)}, the vertical error bars are computed as standard errors from averaging over $M$ realizations. The horizontal error bars are standard errors $\delta E_\alpha=\sqrt{\langle|E^{(j)}_\alpha-E_\alpha|^2\rangle_J/(M-1)}$ that characterize the deviations of $E^{(j)}_\alpha$ from their mean values $E_\alpha$. The error bars in panel {\bf (b)} are defined in the same way. 

Fig.~\ref{fig:ns}. The error bars for the data displayed with the blue squares are computed as standard errors from averaging over realizations. The error bars for the data displayed with red circles are computed as $\sqrt{{\cal O}(E_\alpha)^2 \delta w(E_\alpha)^2 + w(E_\alpha)^2\delta {\cal O}(E_\alpha)^2}$, where we use $w(E_\alpha) = e^{-\beta E_\alpha}/{\cal Z}$. $\delta {\cal O}(E_\alpha)$ is defined as a standard error from averaging ${\cal O}_{\alpha\alpha}^{(j)}$ over realizations. $\delta w(E_\alpha)$ is defined as $\sqrt{(\partial w(E_\alpha)/\partial\beta)^2\delta\beta^2+(\partial w(E_\alpha)/\partial E_\alpha)^2\delta E_\alpha^2}$. $\delta \beta$ stems from the uncertainty $\delta H$ in the total energy $\langle\langle H\rangle\rangle_J$ (calculated as standard errors from averaging over realizations) and the average thermal variance of $H$: $\delta\beta = \delta H/\bigg|\bigg\langle\frac{{\rm Tr} (e^{-\beta H}H^2)}{{\rm Tr}(e^{-\beta H})}-\left[\frac{{\rm Tr} (e^{-\beta H}H)}{{\rm Tr}(e^{-\beta H})}\right]^2\bigg\rangle_J\bigg|$. $\delta E_\alpha$ are the same as defined in Fig.~\ref{fig:ETH}.

Fig.~\ref{fig:YY}. In panels {\bf (a)} and {\bf (b)}, the vertical error bars are computed as standard errors from averaging over the realizations. In panel {\bf (c)}, the vertical error bars are computed as the square root of the covariance of the fit. The horizontal error bars $\delta \beta$ are the same as defined in Fig.~\ref{fig:ns}. In the inset, the vertical error bars are set by $\delta H$, and the horizontal error bars are the same as the vertical ones in the main figure. The length of the black dashes ($E_{\rm in}$) is fixed by the error bars in $\beta$.

\end{document}


\title{Supplemental Material for ``Thermalization in a closed quantum system from randomized dynamics''
}

\author{Nikolay V. Gnezdilov}
\email{nikolay.gnezdilov@dartmouth.edu}
\affiliation{Department of Physics and Astronomy, Dartmouth College, Hanover, New Hampshire 03755, USA}

\author{Andrei I. Pavlov}
\email{andrei.pavlov@kit.edu}
\affiliation{TKM, Karlsruhe Institute of Technology, 76131 Karlsruhe, Germany}

\begin{abstract}
    In this Supplemental Material, we (i) study level statistics of our model; (ii) compare eigenstate occupations and eigenstate expectation values before and after averaging over realizations of the auxiliary interaction; (iii) repeat thermalization from randomized dynamics for different initial states and demonstrate that our thermalization approach is independent of the structure of the initial state; (iv) show equilibration dynamics of a same-time spin-spin correlation function.
\end{abstract}

\maketitle
\setcounter{equation}{0}
\setcounter{figure}{0}
\setcounter{table}{0}
\renewcommand{\theequation}{S\arabic{equation}}
\renewcommand{\thefigure}{S\arabic{figure}}
\renewcommand{\thetable}{S\arabic{table}}

{\bf Level statistics \& energy constraint.} 
Our model, described by the Hamiltonian $H=H_0+{\cal V}^{(j)}$, consists of the target $H_0$ and the auxiliary ${\cal V}^{(j)}$ terms, where each of them is integrable, while their combination is not. The target system is the $1$D transverse-field Ising model. The auxiliary interaction can be mapped onto the random-hopping model using the Jordan-Wigner transform: ${\cal V}^{(j)} = \sum_{i_1>i_2}^N J^{(j)}_{i_1 i_2} c^\dag_{i_1} c_{i_2} + h.c.$, where $c^\dag_i$ and $c_i$ are fermionic creation and annihilation operators at the $i$th site. If considered independently, the target and the auxiliary terms can be represented in terms of free fermions. To avoid integrability, we specifically choose the nearest-neighbor interaction in the target system along $\sigma^y$, so that the target and auxiliary terms cannot be brought to the free-fermion form in the same basis. As a result, adding the auxiliary interaction to the target system makes the system ergodic, which we explicitly demonstrate by considering level statistics of the total Hamiltonian $H=H_0+{\cal V}^{(j)}$.

\begin{figure}[t!!!]
\center 
\includegraphics[width=0.6\columnwidth]{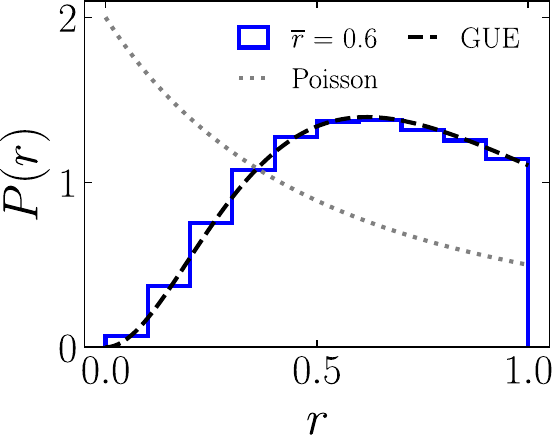}  
\caption{\small  {\bf Probability distribution of the ratio of consecutive level spacings}. The blue histogram shows the probability distribution of $r_\alpha^{(j)}$ sampled from the $200$ realizations of the Hamiltonian $H$ with $N=11$. The dashed black line shows the probability distribution for a system with the level statistics described by the Gaussian unitary ensemble of random matrix theory. The dotted gray line is the probability distribution for a system with Poisson level statistics.
\label{fig:P_WD}
} 
\end{figure}

The many-body energy spectra of ergodic systems follow Wigner–Dyson level statistics of random matrix theory (RMT), in contrast to those of integrable or many-body localized systems, which follow Poisson level statistics~\cite{Berry1977}. To analyze the level statistics of the total Hamiltonian $H$, we use the probability distribution of a ratio $0 \leq r^{(j)}_\alpha \leq 1$  determined from two consecutive level spacings of the many-body spectrum~\cite{Oganesyan2007Localization}: $r^{(j)}_\alpha = {\rm min}\lbrace \delta^{(j)}_{\alpha+1}, \delta^{(j)}_\alpha \rbrace/ {\rm max}\lbrace \delta^{(j)}_{\alpha+1}, \delta^{(j)}_\alpha \rbrace$, where $\delta^{(j)}_\alpha = E^{(j)}_{\alpha+1}-E^{(j)}_\alpha$. 
In Fig.~\ref{fig:P_WD}, we plot the probability distribution of $r^{(j)}_\alpha$ for the $200$ realizations of the Hamiltonian $H$ with $N=11$ spins, excluding the two lowest and two highest energies for every realization. 
The resulting probability distribution in our system is consistent with the probability distribution for the GUE class of RMT, $P(r)= 2(1+r)^2/(1+r+r^2)^4/Z_2$, where $Z_2$ is a normalization constant. The mean value of the gap ratio $\overline{r}$ coincides with the prediction of GUE, $\overline{r}_{\rm GUE} \approx 0.6$~\cite{Atas2013Distribution}. For contrast, in Fig.~\ref{fig:P_WD}, we add the probability distribution $P(r)=2/(1+r)^2$ for the gap ratio of an integrable or many-body localized system~\cite{Oganesyan2007Localization}. 

\begin{figure}[h!!!]
\center 
\includegraphics[width=0.74\columnwidth]{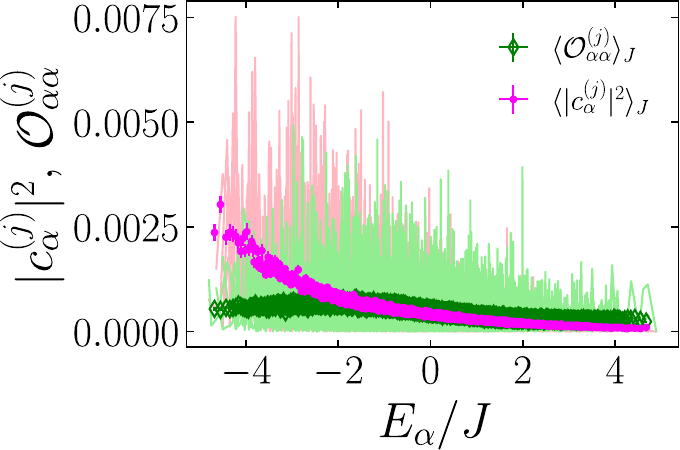}  
\caption{\small  {\bf Eigenstate occupations and eigenstate expectation values}. The green diamonds/magenta dots are the average eigenstate expectation values/eigenstate occupations. The green/magenta lines show every fiftieth realization of eigenstate expectation values/eigenstate occupations before averaging. The error bars are defined the same as in Fig.~1{\bf a} in the main text.
\label{fig:c_and_O}
} 
\end{figure}

Overall, our system belongs to the GUE universality class but respects a global constraint on total energy due to the zero mean of the auxiliary interaction. Indeed, $\langle{\cal V}^{(j)}\rangle_J=0$ ensures that the energy of the initial state $E_{\rm in} = \langle \psi_{\rm in}|H_0|\psi_{\rm in}\rangle$ coincides with the total energy of the system $\langle H\rangle = \langle \psi(t)|H|\psi(t)\rangle=\langle \psi_{\rm in}|H|\psi_{\rm in}\rangle$ {\it on average}: $E_{\rm in}=\langle\langle H\rangle\rangle_J$. 

\begin{figure*}[t!!!]
\center
\includegraphics[width=0.665\columnwidth]{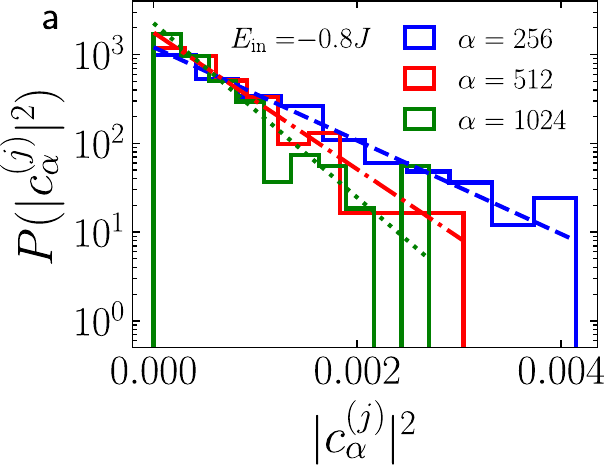}
\includegraphics[width=0.68\columnwidth]{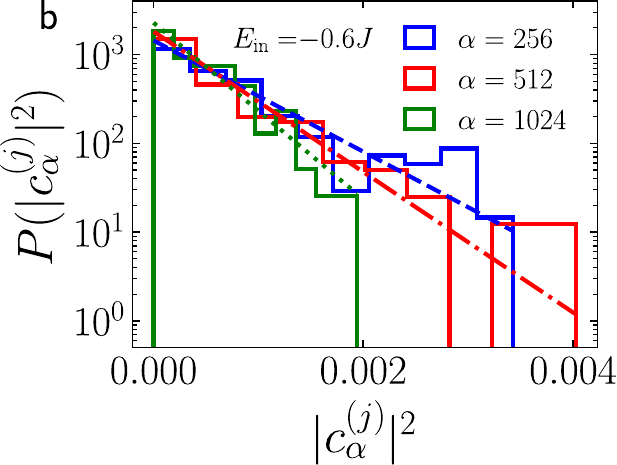}  
\includegraphics[width=0.663\columnwidth]{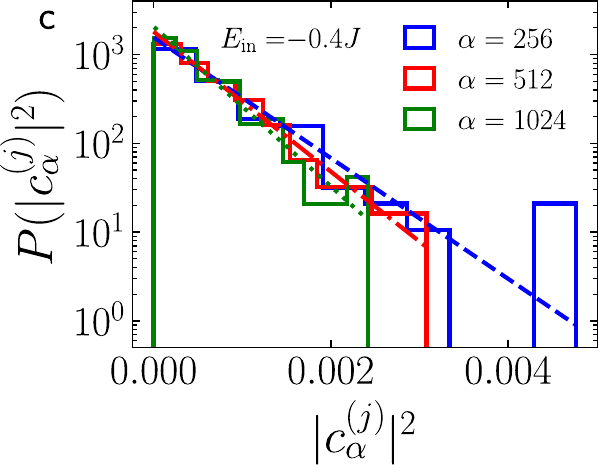} 
\caption{\small {\bf Distributions of $\alpha$th eigenstate occupations for different initial states}. Legends in panels {\bf (a)}--{\bf (c)} are analogous to the legend in Fig.~1{\bf c} in the main text.
\label{fig:PT}
}
\end{figure*}

\begin{figure*}[t!!!]
\center
\includegraphics[width=0.68\columnwidth]{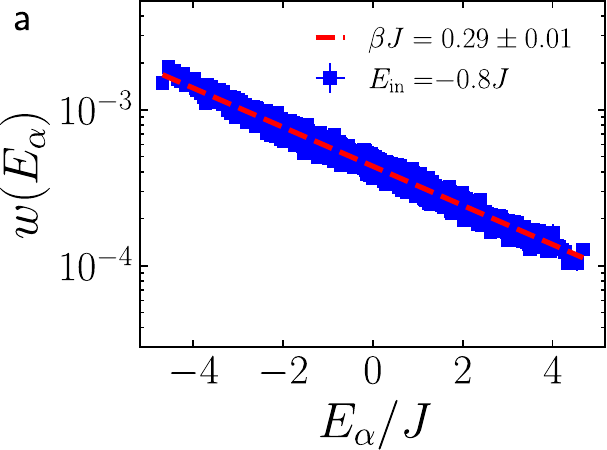}
\includegraphics[width=0.68\columnwidth]{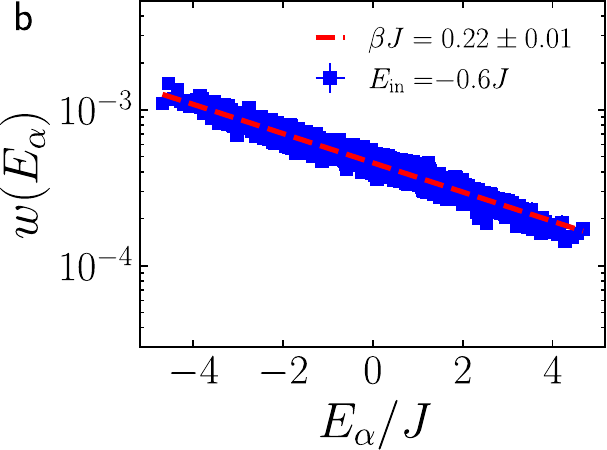}  
\includegraphics[width=0.68\columnwidth]{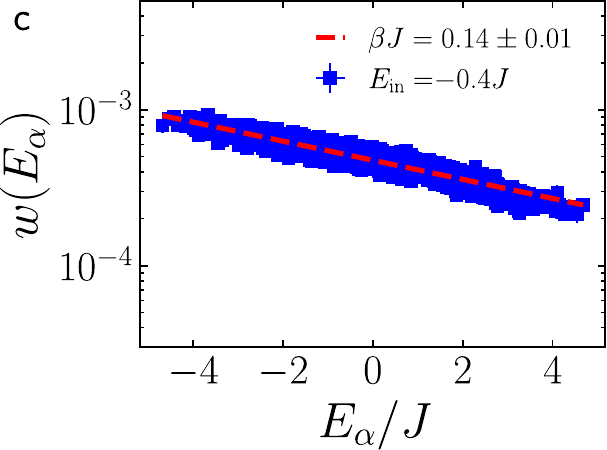} 
\caption{\small {\bf Average eigenstate occupation probabilities for different initial states}. Legends in panels {\bf (a)}--{\bf (c)} are analogous to the legend in Fig.~1{\bf b} in the main text. 
\label{fig:wE}
}
\end{figure*}

\begin{figure*}[t!!!]
\center
\includegraphics[width=0.68\columnwidth]{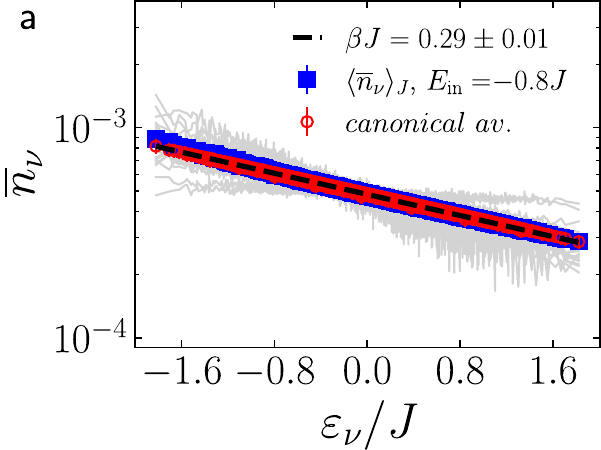}
\includegraphics[width=0.68\columnwidth]{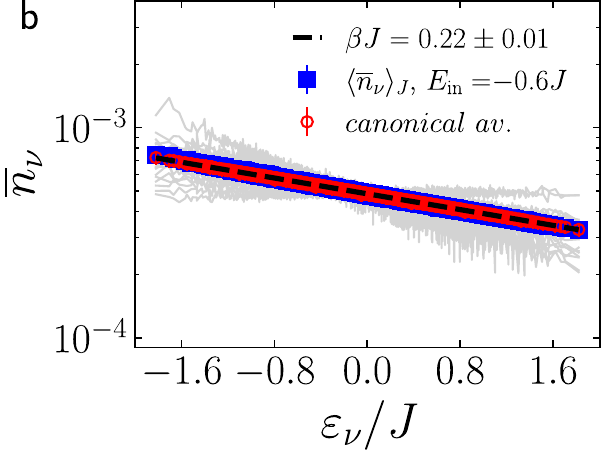}  
\includegraphics[width=0.68\columnwidth]{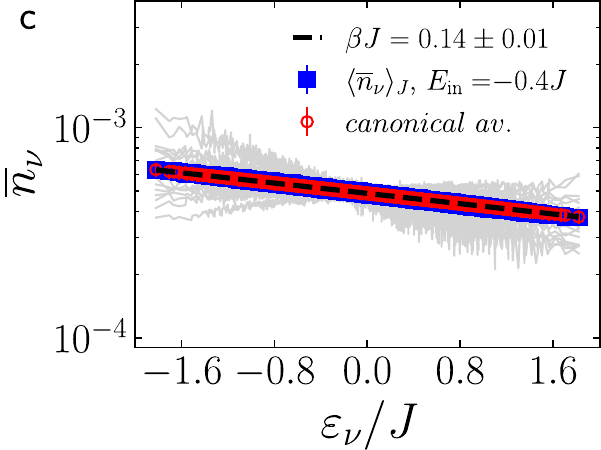} 
\caption{\small {\bf Occupation probabilities of eigenstates of the Ising model for different initial states}.  Legends in panels {\bf (a)}--{\bf (c)} are analogous to the legend in Fig.~2 in the main text.
\label{fig:ns_m}
}
\end{figure*}

\begin{figure*}[t!!!]
\center
\includegraphics[width=0.646\columnwidth]{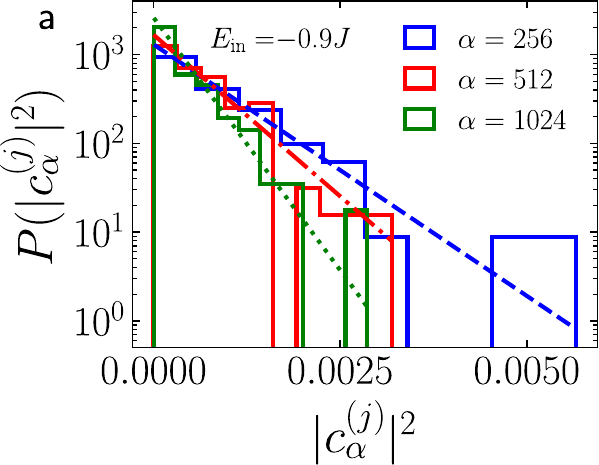}
\includegraphics[width=0.68\columnwidth]{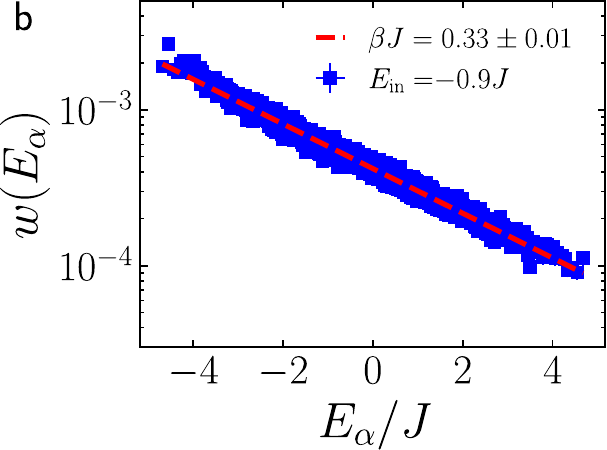}  
\includegraphics[width=0.68\columnwidth]{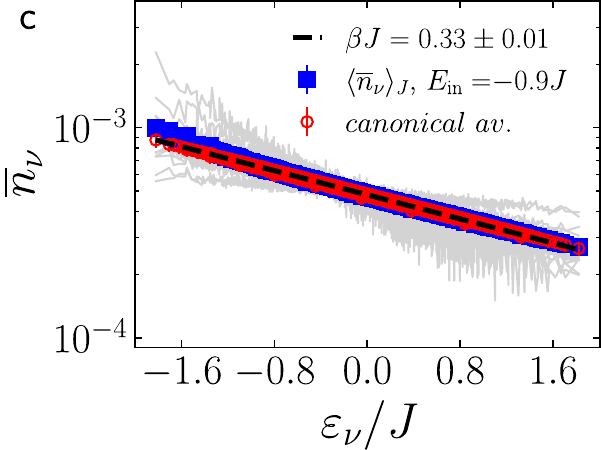} 
\caption{\small {\bf Thermalization from randomized dynamics for the superposition initial state.} Panel ({\bf a}) shows the PT-distributed eigenstate occupations, panel ({\bf b}) shows average eigenstate occupations, and panel ({\bf c}) shows populations of energy levels of the TFIM. Legends and error bars in panels {\bf (a)}--{\bf (c)} are analogous to legends in Figs.~\ref{fig:PT}-\ref{fig:ns_m}, respectively.
\label{fig:sp}
}
\end{figure*}

{\bf The role of averaging over realizations.}
In Fig.~\ref{fig:c_and_O}~\footnote{This figure is a more detailed version of the inset in Fig.~1{\bf a} in the main text}, we show eigenstate expectation values and eigenstate occupations as functions of the many-body energies of $H$. For individual realizations of the auxiliary interaction, the eigenstate expectation values and eigenstate occupations are delocalized across the entire system. We plot the average eigenstate occupations and expectation values as functions of the mean many-body spectrum. The average quantities are substantially smoother, yet remain spread throughout the entire many-body spectrum. This behavior differs from conventional thermalization. In the latter case, eigenstate occupations are narrow in energy, and eigenstate expectation values do not change in a narrow energy window, which, in turn, leads to a microcanonical average for observables~\cite{Rigol2008}. In our case, the average eigenstate occupations take the form of the Gibbs distribution function, leading to a canonical average for any observable, as explained in the main text.

{\bf Different initial states.} 
In Fig.~\ref{fig:PT}, we show distributions of the $\alpha$th eigenstate occupations $|c^{(j)}_{\alpha}|^2=|\langle \psi^{(j)}_\alpha | \psi_{\rm in} \rangle|^2$ for different initial product states with different $E_{\rm in}$. The eigenstate occupations follow the Porter-Thomas (PT) distribution. 
In Fig.~\ref{fig:wE}, we show the average eigenstate occupations $w(E_\alpha)=\langle|\langle \psi^{(j)}_\alpha | \psi_{\rm in} \rangle|^2\rangle_J$ for the same initial states. The eigenstates are on average populated according to the Gibbs distribution with the temperature $\beta^{-1}$ set by $E_{\rm in}$. In Fig.~\ref{fig:ns_m}, we plot occupations of the many-body eigenstates of the Ising model $\overline{n}_\nu = \overline{|\langle\psi(t)|\varphi_\nu\rangle|^2}$. The average occupations are thermal, whereas the occupations calculated for individual realizations (shown as gray lines in Fig.~\ref{fig:ns_m}) deviate from thermal values.
The error bars in Figs.~\ref{fig:wE}, \ref{fig:ns_m} are defined similarly to the error bars in Figs.~1{\bf b}, 2 in the main text, respectively.

\begin{figure}[t!!!]
\center 
\includegraphics[width=0.68\columnwidth]{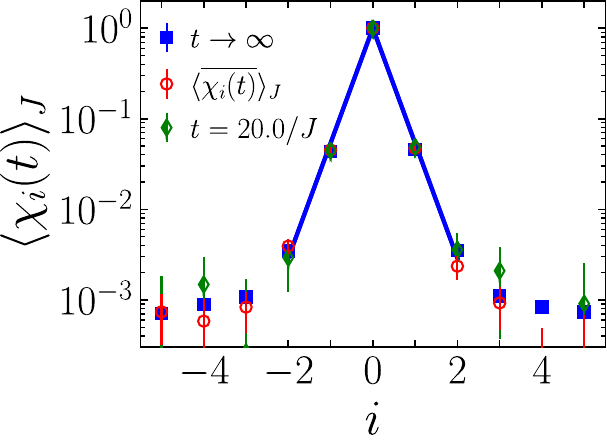}  
\caption{\small  {\bf Spatially resolved same-time spin-spin correlation function averaged over realizations of the randomized Hamiltonian.} The blue squares show the correlation function time-averaged in the long-time limit. The red circles show the correlation function time-averaged over a finite interval. The green diamonds show the correlation function at a fixed time $t = 20/J$. The error bars are standard errors with respect to randomized Hamiltonian realizations of the time-average observable or standard errors at a given time for the $t = 20/J$ plot.  
\label{fig:YY_RT}
} 
\end{figure}

\begin{figure}[t!!!]
\center 
\includegraphics[width=0.64\columnwidth]{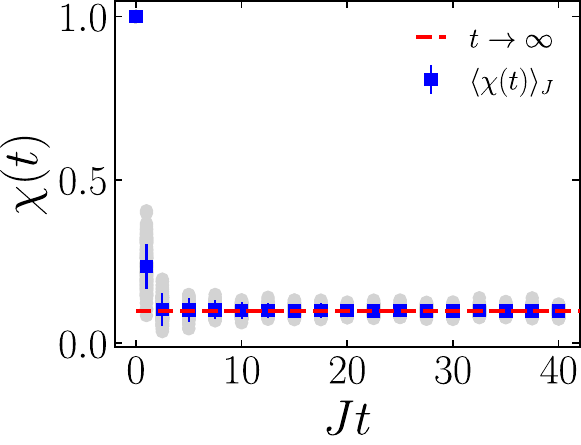}  
\caption{\small  {\bf Same-time spin-spin correlation function as a function of time.} The blue squares show the correlation function averaged over realizations of the randomized dynamics computed at a given time. The corresponding error bars are defined as the sum of squares of errors for the spatially resolved correlation function. The gray circles show the correlation function computed for individual realizations of the randomized dynamics. The dashed red line shows the average correlation function in the infinite-time limit, computed with exact diagonalization.
\label{fig:chi}
} 
\end{figure}

We additionally consider a superposition initial state to demonstrate that our results do not depend on the structure of the initial state. We repeat our thermalization protocol for the superposition initial state and show PT-distributed eigenstate occupations, thermal average eigenstate occupations, and thermally distributed populations of many-body energy levels of the non-perturbed Hamiltonian $H_0$ in Fig.~\ref{fig:sp}. 
The results are qualitatively the same as for the initial product states and depend only on the initial state's energy that sets the resulting temperature.

When comparing the statistics of eigenstate occupations with the PT distribution, we do not use any fits. The straight lines in Figs.~\ref{fig:PT}, \ref{fig:sp}{\bf a} show PT distribution functions, $P(p)=e^{-p/w(E_\alpha)}/w(E_\alpha)$, with the average eigenstate occupations $w(E_\alpha)$ for corresponding many-body energy levels shown in Figs.~\ref{fig:wE}, \ref{fig:sp}{\bf b}. The statistical data gathered for each many-body eigenstate by sampling the auxiliary interaction are in good agreement with the PT distribution for all initial states, with a corresponding first moment given by $w(E_\alpha)$.  

{\bf Real-time dynamics.} We calculated time-averaged observables in the long-time limit using exact diagonalization. Although this is a natural approach to explain our thermalization mechanism, it requires access to the system's eigenstates, which are harder to calculate for larger systems. However, one can use averaging over randomized dynamics to generate thermal observables from the real-time propagation (RTP) by directly evolving the state in Eq.~(4) in the main text. For example, let us consider the spatially resolved same-time spin-spin correlation function,  $\chi_i(t) =\langle \psi(t) | \sigma^y_i\sigma^y_0|\psi(t)\rangle$, analyzed in Fig.~3 in the main text, and compute it directly from the RTP without taking the long-time average and using exact diagonalization. In Fig.~\ref{fig:YY_RT}, we show that the correlation functions computed at a given time ($t=20/J$), or averaged over a finite time interval ($t=[2.5,40]/J$ with a step of $2.5/J$), or computed in the long-time limit, exhibit the same characteristic decay. 
The latter curve is the same as the central correlator for the initial state  $|\psi_{\rm in}\rangle=\otimes_{k=1}^N|\downarrow_y\rangle_k$ in Fig.~3{\bf a},{\bf b} in the main text.

Equilibration to thermal values of average observables does not require considering long times, which may be beneficial for numerical implementations of real-time dynamics, e.g., with the time-dependent variational principle~\cite{Haegeman2011TDVP,Haegeman2016Unifying}. In Fig.~\ref{fig:chi}, we plot a spatially summed correlation function, $\chi(t) =N^{-1}\sum_i\chi_i(t)$, for the same initial state as in Fig.~\ref{fig:YY_RT}. The correlation function equilibrates to the long-time thermal behavior at time $t_{\rm eq}\sim {\cal O}(1)/J$~\cite{Perrin2025Dynamic}. In Fig.~\ref{fig:chi}, we add a time point $t=1/J$ to make a visual reference for the equilibration process.
\linebreak

\bibliography{refs}